# Toward a Push-Scalable Global Internet

Sachin Agarwal
Deutsche Telekom A.G., Laboratories and Technical University of Berlin
Ernst-Reuter-Platz 7
10587 Berlin, Germany
Email: sachin@net.t-labs.tu-berlin.de

***Abstract*—Push message delivery, where a client maintains an "always-on" connection with a server in order to be notified of a (asynchronous) message arrival in real-time, is increasingly being used in Internet services. The key message in this paper is that push message delivery on the World Wide Web is not scalable for servers, intermediate network elements, and battery-operated mobile device clients. We present a measurement analysis of a commercially deployed WWW push email service to highlight some of these issues. Next, we suggest content-based optimization to reduce the always-on connection requirement of push messaging. Our idea is based on exploiting the periodic nature of human-to-human messaging. We show how machine learning can accurately model the times of a day or week when messages are least likely to arrive; and turn off always-on connections these times. We apply our approach to a real email data set and our experiments demonstrate that the number of hours of active always-on connections can be cut by half while still achieving real-time message delivery for up to 90% of all messages.**

## I. Introduction

The Internet is being transformed from a static data repository to a dynamic and real-time information delivery platform. The change is driven by the type of content being uploaded and transmitted through the Internet, particularly via the World Wide Web (WWW). For example, social networking services like Twitter and Facebook serve highly dynamic web pages which may be updated several times every minute. Push email delivery, made popular by the RIM/Blackberry service, is now offered by several web-based email services, including Google's Gmail. The key benefit of push messaging is that users automatically receive the message as soon as it arrives at the server rather than explicitly having to poll the server to check for new messages.

Web servers serve content to user clients, which may be web browsers, email software, or other applications running on users' computers and mobile devices. Most web traffic uses the Hyper Text Transmission Protocol (HTTP) [1] protocol, which runs over the connection-oriented Transmission Control Protocol (TCP) [2]. Real-time message delivery requires an always-on connection from the server to the client because the arrival of a new message on the server is an asynchronous event. This usually means keeping a long-lived TCP connection open between the server and the client over which a message can be "pushed" to the client at anytime it arrives.

Several challenges arise while maintaining long-lived TCP connections on the Internet. First, many network elements, and in particular HTTP proxies through which a majority of users connect, have limited memory and TCP ports that are shared among multiple users. In order to serve more users, HTTP proxies routinely recycle dormant resources. For example, a TCP connection that carries no (HTTP) traffic for a few minutes will be closed by the proxy and this poses a problem for long-lived connections. Second, an active TCP connection may prevent a mobile client from entering power-conserving sleep-modes and thereby reduce its battery life. Third, servers need to be provisioned in order to maintain active TCP connections from large populations of user clients. There is an upper limit (65536) on the number of unique ports available on each server.

In this work we explain some of the commonly used methods to implement push messaging given the above mentioned challenges. We discuss why each method falls short on the scalability front for network elements, clients and servers. We then analyze a real-world push message service for mobile devices to illustrate the scalability bottlenecks of implementing real-time messaging over the global Internet. Next, we suggest an approach that trades the real-time aspect of push-based messaging for scalability and efficiency. By relaxing the condition that *all* messages are delivered to the client in real-time, we drastically reduce the always-on connection requirement. Our approach is based on applying machine learning for identifying temporal periodicity patterns in message arrivals and using this information to mitigate the always-on connection requirement of push messaging. We experiment on a publicly-available email data set to test our approach and demonstrate its benefits.

The paper is organized as follows. In Section II-A we present some related work around push messaging in academia and industry. We explain the common algorithms for WWW push messaging in Section II-B and describe some of the scalability challenges in these algorithms. In Section III we present a measurement study of Google's Android cloud-to-device-messaging (c2dm) push messaging system and report on its underlying always-on TCP connection mechanisms. We then motivate the case for content-based optimization in Section IV based on email arrival time data from the publicly available Enron email data set [3, 4]. Next, we implement our ideas using simple machine-learning algorithms and demonstrate their benefits in Section V. Conclusions and future work directions are discussed in Section VI.

## II. Background

In this section we present pointers to relevant academic and engineering literature about push-based protocols. We then describe different approaches to implementing push messaging protocols between clients and servers on the Internet.

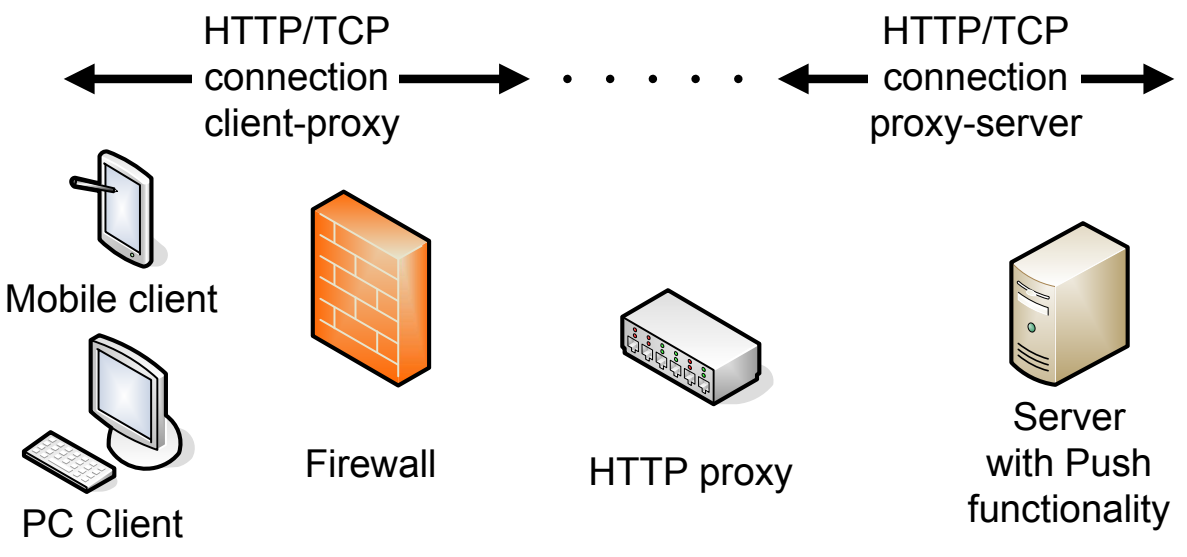

Fig. 1. An HTTP client-server interaction over the Internet. The HTTP proxy and NAT complicate maintaining long-lived TCP connections on arbitrary TCP ports between clients and servers.

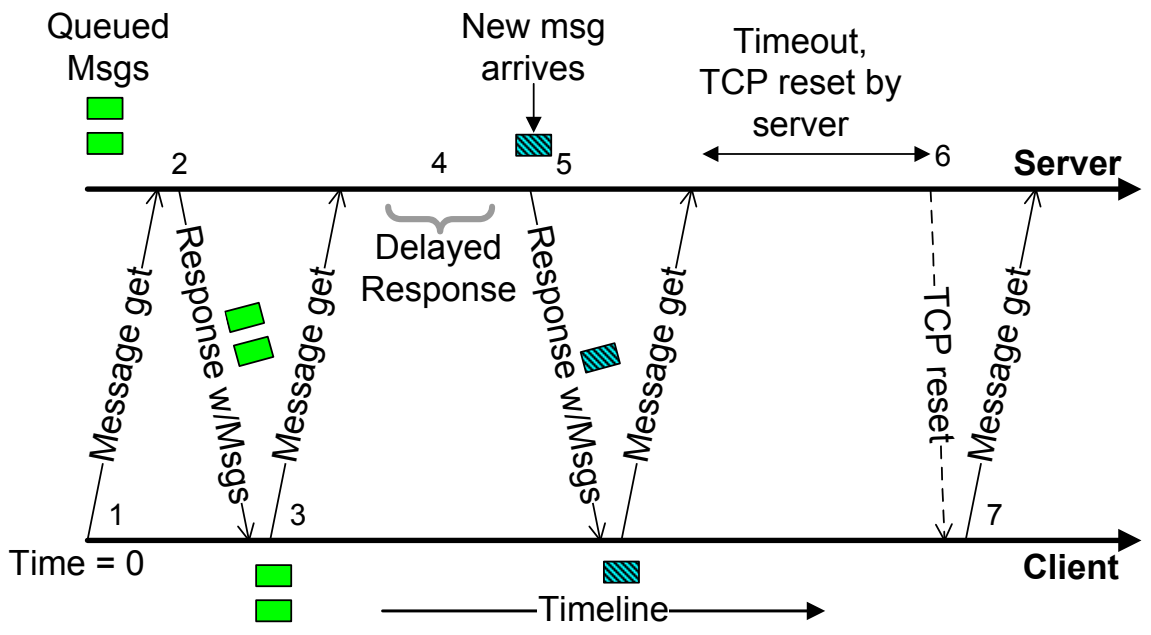

Fig. 2. Time-line of a long polling interaction between a client and server.

### A. *Related Work*

The work on Broadcast Disks [5, 6] used push data dissemination, although not in the Internet/WWW push context. Push-based Internet message delivery was first made popular by the Pointcast messaging service [7] in 1996. The first academic treatment of Internet push was in [8], where the authors presented an overview of the push techniques and challenges of Internet/WWW push.

The idea of pushing information to clients has been an enabler for publish-subscribe mechanisms [9, 10]; RSS (real simple syndication), the canonical publish-subscribe example, employs a form of short polling (Section II-B) by clients in order to download content updates. Web-based push messaging took flight with the advent of AJAX [11] technology which allows content dynamic updates within a web page without reloading the whole web page. Studies such as [12] have pointed toward a noticeable shift in HTTP network traffic properties due to the automatic page reloading characteristics of AJAX. Our work specifically looks at how push messaging is implemented over HTTP and then suggests methods to reduce the overhead of message push.

### B. *Web Client-Server Architecture and Push Protocols*

Fig. 1 depicts a typical WWW client-server connection. User clients are usually located behind a NAT (Network Address Translation) device or firewall, which may be a home router or a corporate firewall with a HTTP proxy that disallows non-HTTP(S) TCP ports and protocols for security reasons. Most wireless cellular data services for mobile devices invariably use HTTP proxies and NAT devices. Proxies serve as content caches and are used to enforce content policing and user monitoring policies. HTTP proxies act as intermediaries between a client and server during HTTP data exchange but they are not configured to maintain long-lived TCP connections between the client and server.

HTTP proxies may buffer HTTP data and thereby introduce unbounded latency for WWW push messaging. This can be circumvented using HTTPS [13] connections that are transparently tunneled (without buffering) by most proxies. The Android c2dm service discussed in Section III sets up HTTPS connections in order to traverse HTTP proxies transparently. But even HTTPS cannot prevent proxies from closing seemingly dormant connections and protocols have to be engineered to overcome this limitation. Moreover, HTTPS increases the protocol and computational overhead as compared to plain HTTP on servers and clients.

We now briefly describe some common algorithms for implementing push message delivery [14].

**Short polling** is commonly used to implement push message delivery. Each client periodically polls the server for new messages. The advantage of short polling is that it does not require long-lived TCP connections. The whole transaction of connecting to the server and retrieving any waiting messages can be completed in one HTTP interaction, making short polling a stateless alternative to long-lived TCP connections. The polling period $T$ can be chosen according to application requirements. For a time interval $(0, t)$, a short-polling client will connect to the server $\frac{t}{T}$ times. If message arrival times are independent and uniformly distributed in $(0, t)$ then a message will suffer an expected delay of $\frac{T}{2}$ before it is delivered to the client. Further, if $n$ clients connect to the server then the server receives $n\frac{t}{T}$ TCP connection requests in the time interval $(0, t)$.

**Long polling** Fig. 2 shows the interaction between a client and server during long polling. At time 0 the client sends a (HTTP) request to the server for new messages - labeled (1). The TCP connection carrying the HTTP request is kept open (while the client waits for a response). The server sends back a response containing messages that were queued up before the client came on-line (2). After receiving the messages the client closes the TCP connection, and immediately sends another request for new messages to the server (3). But no new messages have arrived at the server and it delays sending a response (4) to the client until the time a new message is available, at which time (5) it simply transfers the new message to the client. Again, the HTTP transaction is now complete and the TCP connection is closed. The client restarts the long polling process immediately thereafter.

Ideally, if $m$ messages arrive in the time interval $(0, t)$ then the client sends at most $m$ polling requests to the server. In reality if the server response is delayed beyond a few minutes then the open TCP connection is reset (6) and the message request is re-sent by the client (7). Long polling algorithms are employed in the "Comet" [15] APIs and in the Android c2dm service discussed in Section III.

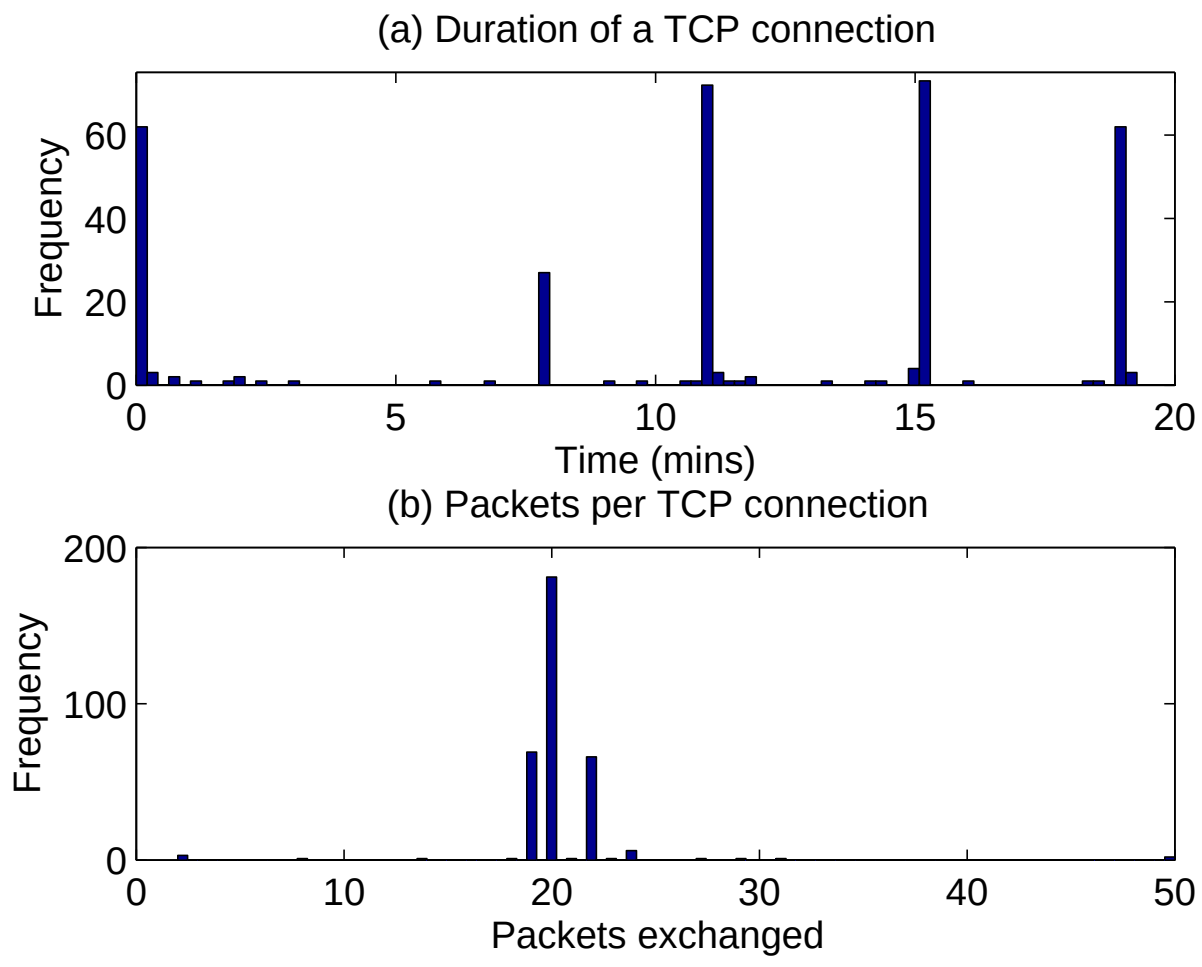

Fig. 3. Histograms of (a) longevity of TCP connections and (b) packets exchanged during each TCP connection.

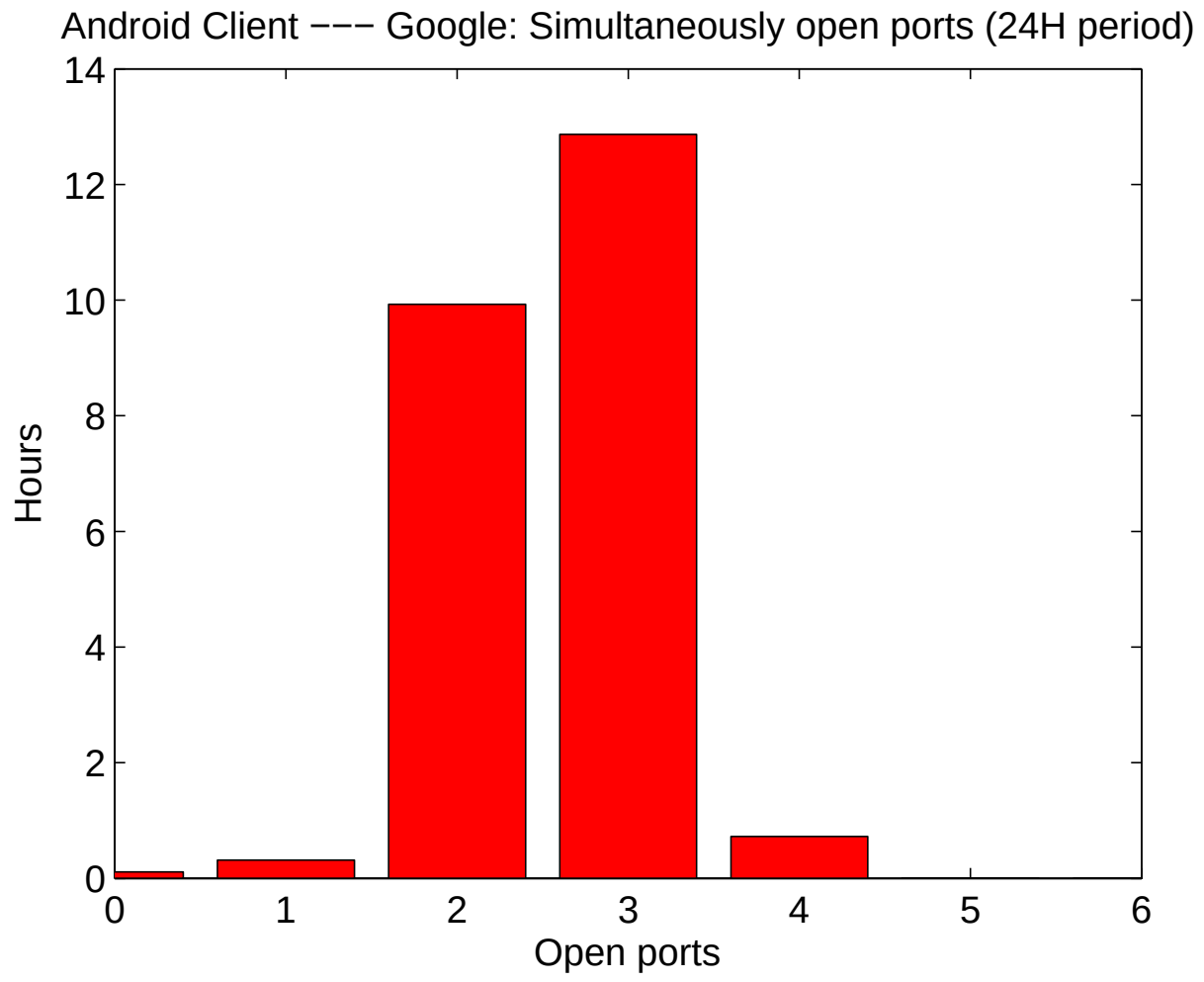

Fig. 4. Number of simultaneously open ports between the Android client and the c2dm server(s)

**Websockets** are part of the emerging HTML 5 [16] specification. The idea behind websockets is to use one long-lived TCP connection to multiplex several back-to-back HTML poll requests and pushed messages from a server to the client. In contrast to long polling, websockets do not "close" the TCP connection after a message is pushed from the server to the client. Nevertheless, the requirement for long-lived and always-on connections remains unchanged and TCP connections have to be restarted every few minutes [17].

## III. Case-study: Android Push Messaging

Large web service providers, notably Google, have constructed application layer push protocols over HTTP(S) to implement push messaging. In Google's case, all push messages destined for a particular user (email, calendar updates, etc.) are delivered to a push server, which in turn delivers them to the user. This push infrastructure is also available to third-party services via the c2dm API for the Android 2.2. mobile operating system. Aggregating all push connections via a single push server reduces the number of push connections to one per device (instead of one per service).

We now present the interaction between Android clients and Google's servers. Our measurement consisted of capturing network traces of exchanges between Google's servers and a client running the Android 2.2. 'Froyo' operating system. We used the Android Gmail application as the target push-based application. The Wireshark network protocol analyzer was used to capture TCP conversations between Google servers and the Android client over a 24 hour period. All communication between the Android client and the Google servers occurred over a secure HTTPS channel. Therefore we are unable to report specific packet-level protocol details. Instead, we extracted several connection-level statistics that are presented below.

Android c2dm uses long polling to implement push messaging. Fig. 3 shows histograms of the longevity of TCP connections and the number of (IP) packets exchanged during each TCP connection. A total of 335 TCP connections were recorded over a 24 hour period with a mean longevity of 663 seconds. Almost all TCP connections lasted for about less than a second or 7, 11, 15 or 19 minutes. Almost all TCP connections exchanged 19, 20 or 22 packets. The regularity in both the histograms clearly indicates that the TCP connections are being opened and closed via algorithms running on the server and client (and not through random network events).

Fig. 4 shows the number of simultaneous ports (TCP connections) open between the Android client and the c2dm server(s). It is interesting that 2-3 ports were open at most times; this could be for introducing redundancy (in case a TCP connection becomes non-responsive) or to avoid the slow-start latency issue with TCP by opening multiple parallel TCP connections.

The Android c2dm algorithms kept sustained active connections between Google servers and the Android client for almost the entire duration of the experiments. But only 8 emails arrived in the 24 hour period. The scalability of this approach with increasing number of clients is suspect; the client needs to keep multiple TCP ports open, the server needs to run the long-polling algorithm for ever client, and the intermediate proxies need to tunnel multiple HTTPS connections per client on a sustained basis.

## IV. Content-Based Optimization

In Section II-B we explained common techniques for implementing push messaging on the WWW and also described the weaknesses of each approach. The Android c2dm case study in Section III illustrated a real implementation of long polling and highlighted the engineering challenges of maintaining an always-on connection between the client and the push server. The scalability of any WWW push protocol is limited by the necessity of long-lived TCP connections. Network components like HTTP proxies cannot maintain long-lived stateful TCP connections (expected grow linearly with the number of users and the number of different push servers to which a client is connected). The present workaround is to repeatedly create and tear down TCP connections and to use HTTPS to tunnel push-traffic.

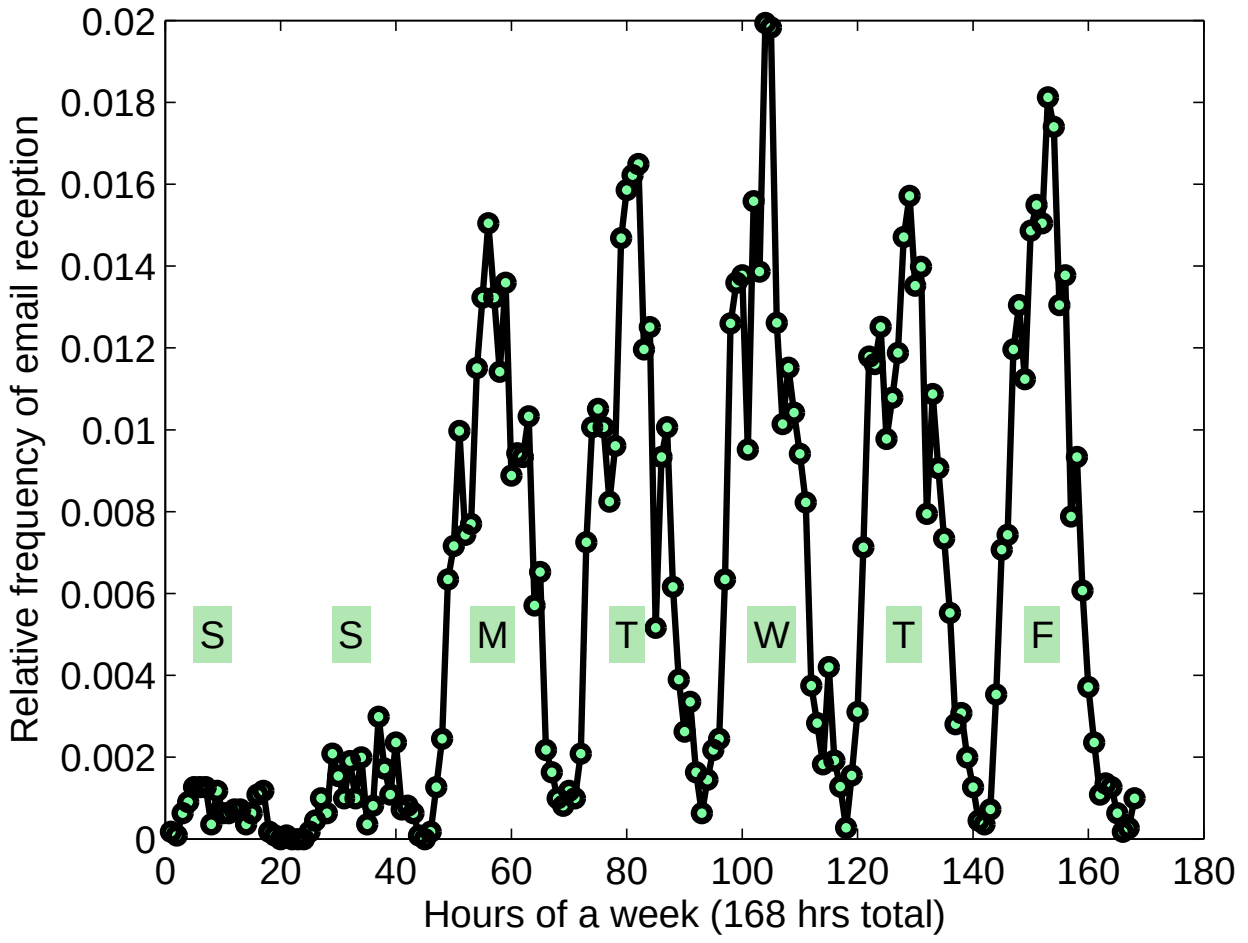

Fig. 5. Periodic nature of email reception - one user's weekly email reception (averaged over 110 weeks). Days-of-the-week are also marked.

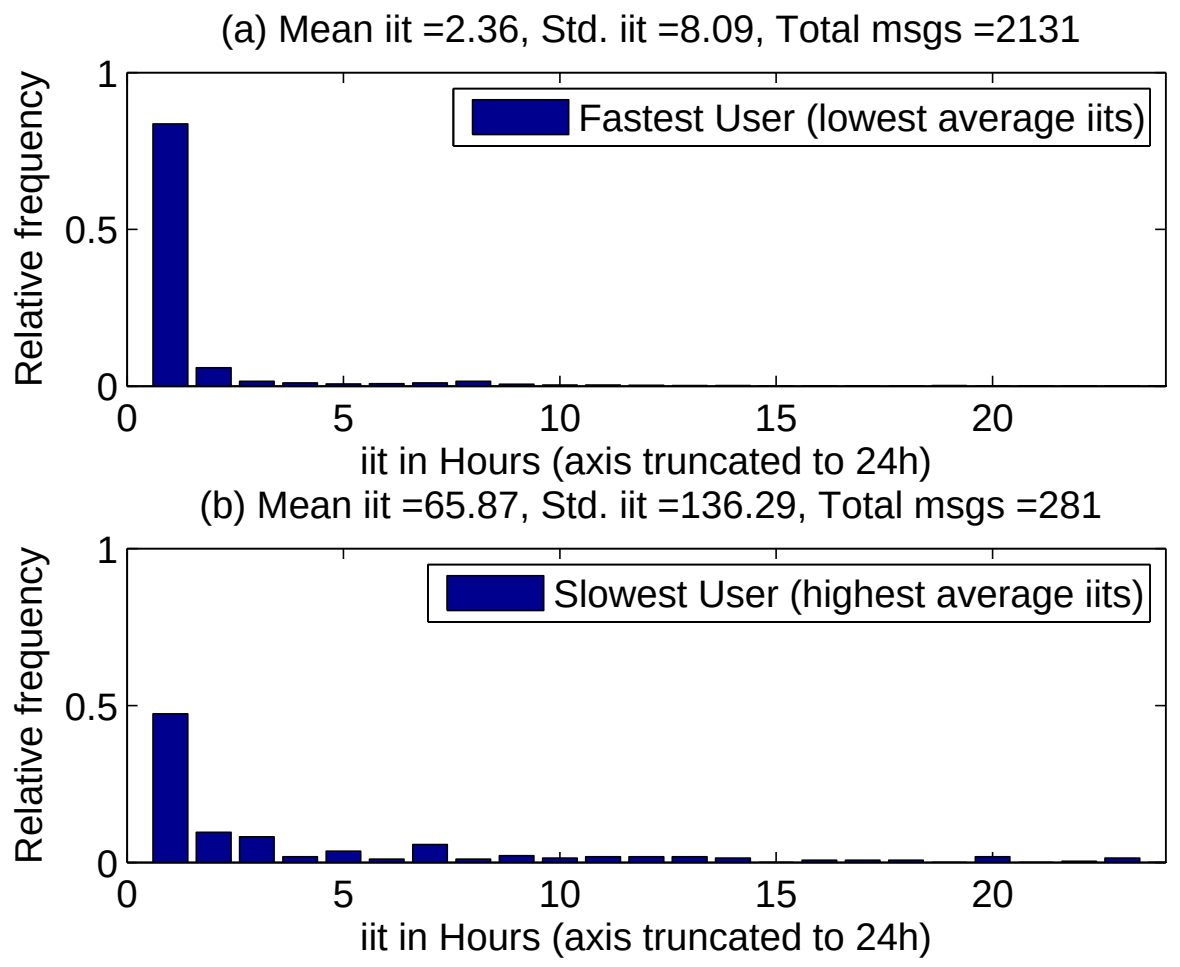

Fig. 6. Relative frequency of inter-arrival times (iits) for (a) the fastest user and for (b) the slowest user. Time is reported in hours. The means, averages, and number of messages is also reported on the plots.

But this does not address the core scalability issue of maintaining always-on connections. Since it is unpractical to hope that the network equipment will be updated to handle the adoption of push messaging on the Internet globally, an algorithmic approach to reducing the need for always-on TCP connections can enable the present infrastructure to support many more users by making more efficient use of resources.

Our key idea is to reduce the time for which a client is connected to the server by analyzing past message arrival patterns. We show how past message reception times are good predictors of future message arrival times due to the periodic and repetitive nature of human communication. Fig. 5 shows the average number of email arrivals in a user's office email inbox during different times of the week (the data-set for this example is described in Section V-A). Unsurprisingly, there is an innate periodicity in message arrivals according to the time of the week.

Fig. 6 shows histograms of email inter-arrival times (iits) of the fastest and slowest user (in terms of average iits) among the 150 users contained in the data set. The plot shows that there are substantial time gaps between email arrivals. Moreover, the mean and standard deviation values of the iits suggest that the iits are not exponentially distributed and therefore the email arrivals process is not Poisson (An exponential distribution with rate parameter $\lambda$ has mean $\frac{1}{\lambda}$ and corresponding standard deviation $\frac{1}{\lambda}$, but this is certainly not the case here). Intuitively this means that continuous TCP connections between the client and push server yield little benefit for many hours in an average week because message arrivals are not "spread out"; instead they are clustered in certain hours. Closing push connections during anticipated periods of low activity would improve the overall scalability of push messaging.

It would be impossible to avoid some messages from being queued at the server should they arrive at the server when no push connection is active between the client and server. Our task is to design machine learning algorithms that minimize possibility of this occurrence, or equivalently, maximize the number of messages that are delivered in real-time (meaning, an active push connection exists between the client and server when the message arrives) given a fixed quota of push connections per unit time. Our learning algorithms should tell us when to turn on or turn off push connections.

In addition to past arrival times, there may be other information in past messages that can assist in coming up with the best times to maintain a push connection. For example, a user may want to unequally weigh the importance of messages from certain people or with certain semantic content. Information about message size, its spam score, etc. can also provide additional information. In this work we consider message arrival times only. Our goal is to demonstrate the possible gains of content-based optimization rather than achieving the best and most optimal push connection "on-times" for a specific user.

## V. Experiments

In this Section we apply machine learning techniques to decide when the always-on connection can be turned off. We test our approach on the publicly available Enron email data set, described below.

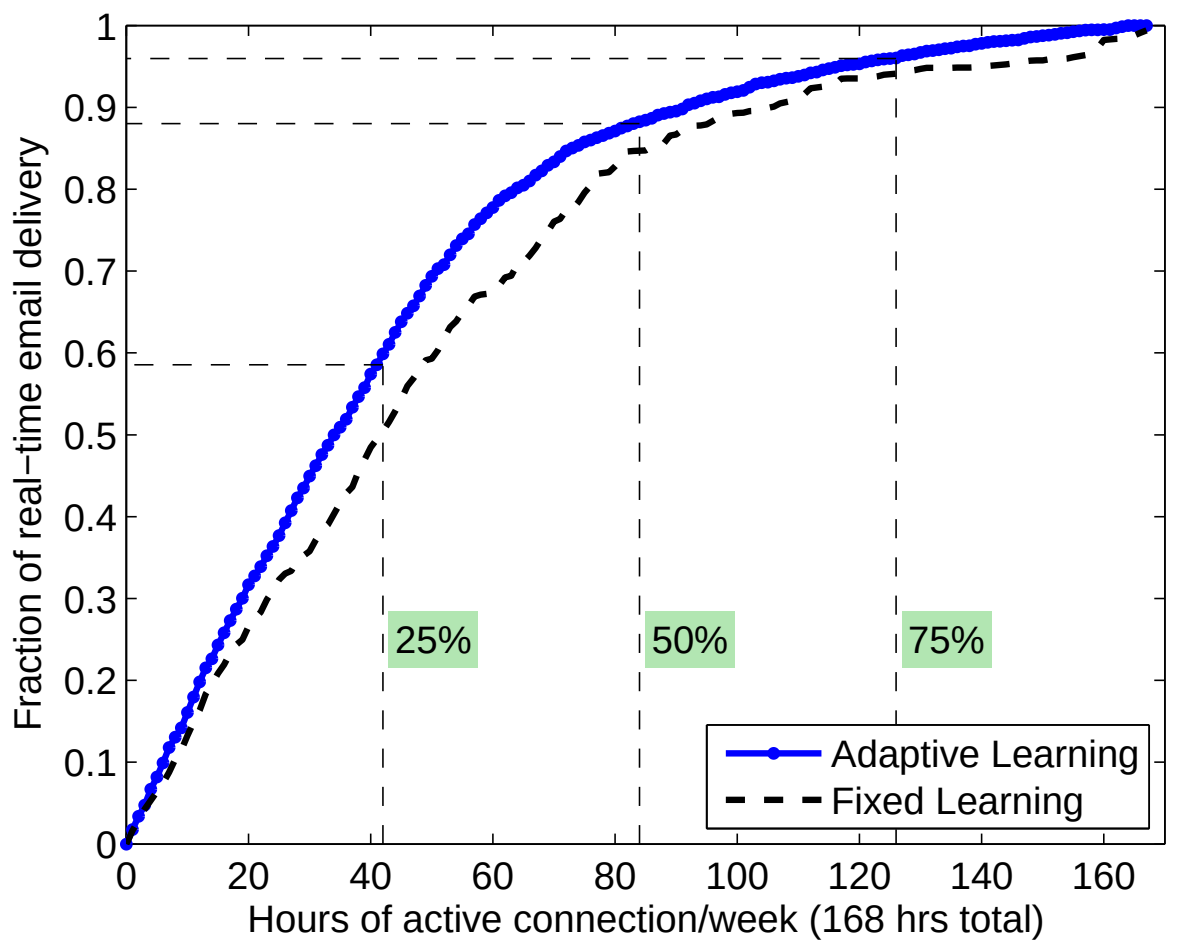

Fig. 7. Single user, $\alpha = 0.9$ for adaptive learning

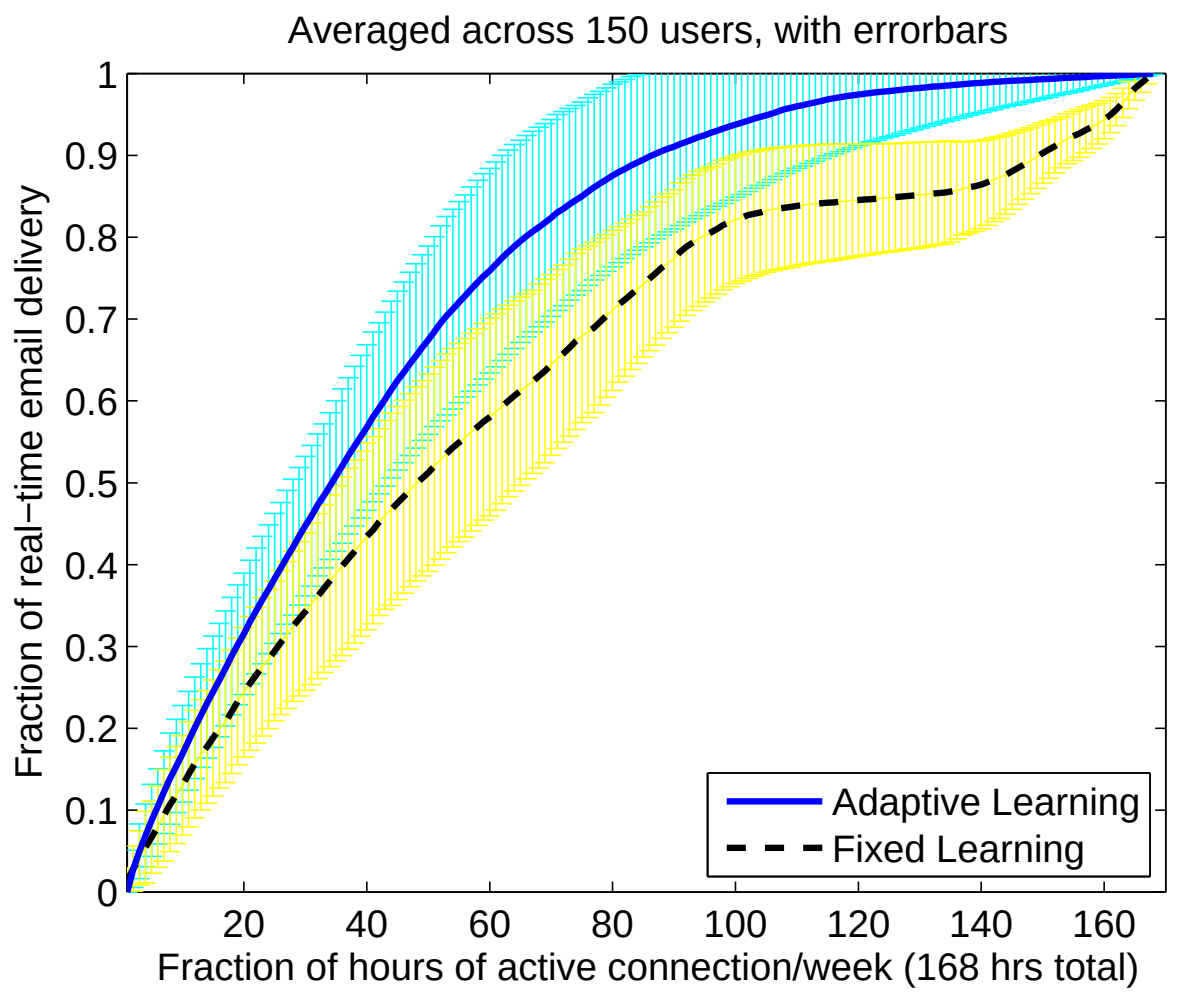

Fig. 8. 150 users, $\alpha = 0.9$ for adaptive learning

## A. Enron Email Data Set

The Enron email data set [3, 4], publicly released by the US Federal Energy Regulatory Commission, contains approximately $500,000$ email messages of 150 senior Enron employees (with email addresses ending in '@enron.com') over a span of about 4 years. We extracted the sending time of each email from its header and used this information to build lists of email-reception times for each employee. We assumed that the network transmission time of each email is negligible (so the sending time was the receiving time at the recipient's inbox in the corporate email server); this assumption is reasonable for the data-set because most of the emails were sent by Enron employees to Enron employees over the corporate Enron network, resulting in near instantaneous email delivery at the recipient's inbox in the corporate email server.

Our learning algorithms are based on the assumption that past email arrival rates at a certain time of the week are good predictors of email arrival rates at the same time in a future week. We use the past email arrival rates in order to "rank" the benefit of keeping push connections alive (or not) at a certain time (compared to other times). Given the fixed cost of setting up a TCP connection we assume that a TCP connection is kept alive for a minimum time (at least one hour in our experiments). We binned the email arrival times for each user into 1-hour bins over each 1 week period (168 hours) in order to capture the daily and weekly email arrival periodicity. After ranking each hour of the week, our algorithm simply picks up the top-$k$ ranked hours ($0 \leq k \leq 168$) as the hours of the week when the push connection is to be kept active.

## B. Learning Approaches

We now describe two simple machine learning approaches for ranking the hours of the week. We emphasize that these learning approaches are rather simplistic and better ranking should be possible with more advanced machine learning algorithms, for example, using neural networks. The objective of our experiments here is to show the first order savings possible through even simple learning techniques.

**Fixed Learning** In fixed learning we use a pre-determined fraction of past email arrival times in order to learn the relative email arrival rates for each hour-of-the-week for a user. For example, if the data set contains 100 weeks of email arrival data for a user and we choose the pre-determined fraction to be $0.1$, then we use the first 10 weeks' email arrival data to generate a 168-element vector (with the index corresponding to the hour-of-the week) containing the relative frequency of email arrival for each hour. We use this relative frequency to rank hours-of-the-week according to email arrival activity. The other part of the arrival data (the last 94 weeks) is used to test the learning algorithm.

**Adaptive Learning** In this approach the hour ranking is updated for every week tested based on all previous weeks (and not just a pre-determined fraction of the total arrival-time data available). More recent data is given a higher weight than older data. In particular, if $F_i$ is the 168-element vector of relative frequencies of email arrival for weeks 0 through $i-1$ and $f_i$ is the vector of relative frequency arrivals for week $i$ then,

$$F_{i+1} = \alpha F_i + (1-\alpha) f_i \tag{1}$$

Where $0 \leq \alpha \leq 1$ determines the relative importance of more recent email arrival times as compared to earlier email arrival times.

## C. Results

We now present results that demonstrate the trade-off between real-time message delivery and adopting a content-based optimization approach on when to keep the push connection alive as opposed to the always-on approach of keeping push connections alive at all times. Fig. 7 shows results for the fastest user (highest average number of message arrivals per unit time) among the $150$ users. The figure shows that about 90% of messages can be delivered in real-time by keeping the push-connection alive for only $50\%$ of the time. It is interesting to note that the difference between the fixed and adaptive learning algorithms is marginal in this case because $10\%$ of the data used for fixed learning is enough to generate a reasonable model for this user. Still, adaptive learning tracks changes in email arrival patterns over time and outperforms fixed learning consistently.

Fig. 8 has the same axes as the previous figure but this figure is plotted for all the $150$ users in the Enron email data set. Error bars are also plotted. Here the limitation of using only $10\%$ data in the fixed learning algorithm is apparent. In particular, for users with sparse arrival data (only few message arrivals), the fixed learning algorithm is not able to capture the periodic weekly variation based on just $10\%$ of the arrival data. In particular, the algorithm does not adequately learn that some hours-of-the-week have very low email arrival rates. This results in a dip in the learning accuracy at about $100$ hours. This corresponds to $48$ weekend hours plus about $4$ hours per weekday during the dead of night ($168-48-4\times 5=100$). In contrast, the adaptive algorithm does not suffer this limitation.

## VI. Conclusion and Future Work

In this paper we investigated the scalability challenge in real-time push message delivery on the WWW. We described the lack of scalability with respect to network limitations (tied up IP addresses and HTTP proxy resources), the battery life of mobile devices (always-on TCP connections that are repeatedly setup and torn-down), and increased CPU processing on mobile devices (HTTPS connections to transparently connect through HTTP proxies in the network). Next, we analyzed the Android (Gmail) push-based message delivery service in order to exemplify some of these scalability issues. Finally, we showed how content-based optimization can be used to remedy the always-on connection requirement, a key scalability limiter, at the price of a few messages not being delivered in real-time.

In particular, we showed how simple machine learning approaches can detect temporal periodicity in past message arrival frequencies and then use this information to turn off the push connection when the probability of receiving a message is low. We demonstrated this approach by applying it to the publicly-available Enron email data set. Our results show that the always-on connections can be cut in half while still achieving real-time message delivery for about 90% of all messages.

Messages that arrive during the times when there is no active push connection between the client and the server will be queued up at the server and their delivery will be delayed. But because the temporal periodicity extracted by the machine learning algorithms roughly corresponds to periods of low human activity (e.g. nights and weekends), users may not be averse to adopting this approach. Messages marked urgent might also be delayed, but mobile devices offer other asynchronous signaling mechanisms (e.g. short messaging service - SMS) to remedy this situation. For example, the server could send an SMS message to the client in case a message marked urgent arrives at a time when there is no active push connection.

This work can be extended in multiple ways. Other message arrival data sets can be analyzed for periodic patterns (e.g. social network updates from services such as Twitter or Facebook). More advanced machine-learning algorithms may yield higher savings (e.g. neural networks) as compared to the approaches investigated in this paper. Finally, more information can be extracted from messages than just arrival information. For example, a user may want to the machine learning algorithm to unequally weigh the importance of a message based on the sender, content, etc.

## References

[1] R. Fielding, J. Gettys, J.Mogul, H. Frystyk, L. Masinter, P. Leach, and T. Berners-Lee. RFC: Hypertext transfer protocol – http/1.1. [Online]. Available: http://www.w3.org/Protocols/rfc2616/rfc2616.html

[2] U. I. S. Institute. RFC 793 - Transmission Control Protocol. [Online]. Available: http://www.faqs.org/rfcs/rfc793.html

[3] W. W. C. (CMU). Enron Email Dataset. [Online]. Available: http://www.cs.cmu.edu/ enron/

[4] A. Fiore and J. Heer. UC Berkeley Enron Email Analysis (database representation]. [Online]. Available: http://bailando.sims.berkeley.edu/enron_email.html

[5] S. Acharya, R. Alonso, M. Franklin, and S. Zdonik, "Broadcast disks: data management for asymmetric communication environments," in *Proceedings of the 1995 ACM SIGMOD international conference on Management of data*, ser. SIGMOD '95. New York, NY, USA: ACM, 1995, pp. 199–210. [Online]. Available: http://doi.acm.org/10.1145/223784.223816

[6] S. Acharya, M. Franklin, and S. Zdonik, "Balancing push and pull for data broadcast," vol. 26. New York, NY, USA: ACM, June 1997, pp. 183–194. [Online]. Available: http://doi.acm.org/10.1145/253262.253293

[7] Wikipedia. Pointcast.com. [Online]. Available: http://en.wikipedia.org/wiki/PointCast_(dotcom)

[8] M. Franklin and S. Zdonik, ""data in your face": push technology in perspective," in *Proceedings of the 1998 ACM SIGMOD international conference on Management of data*, ser. SIGMOD '98. New York, NY, USA: ACM, 1998, pp. 516–519. [Online]. Available: http://doi.acm.org/10.1145/276304.276360

[9] K. Birman and T. Joseph, "Exploiting virtual synchrony in distributed systems," in *Proceedings of the eleventh ACM Symposium on Operating systems principles*, ser. SOSP '87. New York, NY, USA: ACM, 1987, pp. 123–138. [Online]. Available: http://doi.acm.org/10.1145/41457.37515

[10] P. T. Eugster, P. A. Felber, R. Guerraoui, and A.-M. Kermarrec, "The many faces of publish/subscribe," *ACM Comput. Surv.*, vol. 35, pp. 114–131, June 2003. [Online]. Available: http://doi.acm.org/10.1145/857076.857078

[11] Wikipedia. AJAX programming. [Online]. Available: http://en.wikipedia.org/wiki/Ajax_(programming)

[12] F. Schneider, S. Agarwal, T. Alpcan, and A. Feldmann, "The new web: characterizing ajax traffic," in *Proceedings of the 9th international conference on Passive and active network measurement*, ser. PAM'08. Berlin, Heidelberg: Springer-Verlag, 2008, pp. 31–40. [Online]. Available: http://portal.acm.org/citation.cfm?id=1791949.1791955

[13] E. Rescorla. RFC: Http over tls. [Online]. Available: http://www.faqs.org/rfcs/rfc2818.html

[14] Wikipedia. Push technology. [Online]. Available: http://en.wikipedia.org/wiki/Push_technology

[15] ——. Comet programming. [Online]. Available: http://en.wikipedia.org/wiki/Comet_(programming)

[16] W3C. HTML 5 specification. [Online]. Available: http://www.w3.org/TR/html5/

[17] P. Lubbers. How html5 web sockets interact with proxy servers. [Online]. Available: http://www.infoq.com/articles/Web-Sockets-Proxy-Servers